\documentclass[letterpaper, 10 pt, conference]{ieeeconf}

\IEEEoverridecommandlockouts
\usepackage[T1]{fontenc}
\usepackage{lmodern}
\usepackage{amsmath}
\usepackage{mathptmx}                 % Times text and math, as IEEE requires
\usepackage{amssymb}
\usepackage{amsthm}
\usepackage{bm}
\usepackage{graphicx}
\usepackage{microtype}
\graphicspath{{figs/}{./}}
\usepackage{booktabs}
\usepackage{cite}
\usepackage{tikz}
\usetikzlibrary{arrows.meta,positioning,calc,decorations.pathreplacing}
\usepackage{hyperref}
\hypersetup{
    colorlinks=true,
    linkcolor=blue,
    citecolor=blue,
    urlcolor=blue,
    filecolor=blue
}

\definecolor{cA}{HTML}{C0392B}   % event e / suppression
\definecolor{cB}{HTML}{1B4F9C}   % event f / efforts
\definecolor{cC}{HTML}{1E8449}   % preservation / payoffs
\definecolor{cE}{HTML}{B9770E}   % scores
\definecolor{cG}{HTML}{34495E}   % neutral ink
\definecolor{ink}{HTML}{263746}
\definecolor{blueA}{HTML}{2878B5}
\definecolor{blueL}{HTML}{E8F2FA}
\definecolor{redA}{HTML}{C6534B}
\definecolor{redL}{HTML}{FBECEA}
\definecolor{goldA}{HTML}{D99A27}
\definecolor{goldL}{HTML}{FFF4D9}
\definecolor{tealA}{HTML}{218C83}
\definecolor{tealL}{HTML}{E5F4F1}
\definecolor{greenA}{HTML}{4C956C}
\definecolor{greenL}{HTML}{EAF4ED}

\makeatletter
\def\thm@space@setup{\thm@preskip=2.5pt \thm@postskip=2.5pt}
\makeatother
\theoremstyle{plain}
\newtheorem{theorem}{Theorem}
\newtheorem{lemma}{Lemma}
\newtheorem{proposition}{Proposition}
\newtheorem{corollary}{Corollary}
\theoremstyle{definition}
\newtheorem{definition}{Definition}
\newtheorem{remark}{Remark}

\newcommand{\R}{\mathbb{R}}
\newcommand{\E}{\mathcal{E}}
\newcommand{\NN}{\mathcal{N}}

\title{\LARGE \bf Competing for Collective Memory: Strategic Visibility under Limited Attention\\
}

\author{Gurkirat~Wadhwa$^{1}$ and Riya~Sultana$^{1}$%
\thanks{$^{1}$Industrial Engineering and Operations Research,
Indian Institute of Technology Bombay, Mumbai 400076, India.
Emails:
\href{mailto:gurkirat@iitb.ac.in}{gurkirat@iitb.ac.in};
\href{mailto:riya_sultana@iitb.ac.in}{riya\_sultana@iitb.ac.in}.}%
}
\begin{document}
\maketitle
\thispagestyle{empty}
\pagestyle{empty}

%==============================================================================
\begin{abstract}
We study how agents compete over which past events stay visible on a platform with limited display capacity. Agents spend budgeted resources to preserve favourable events or suppress unfavourable ones, while the platform divides a fixed capacity across all events by a softmax rule. Because the shares have one common denominator, an agent's effort also moves events it does not target. We establish mixed- and pure-strategy equilibrium existence. For quadratic effort costs we derive an explicit curvature bound, independent of the number of events, giving equilibrium uniqueness and a step-size range for geometric convergence of synchronous projected-gradient play. We also give an exact criterion for visibility-ranking reversal, show that events receiving no effort are rescaled by one common factor, and derive a sufficient memory-share guarantee for a prescribed coordinated preservation policy. Numerical examples over a finite design grid separate the mismatch of the no-effort allocation from the displacement strategic effort adds, including instances with overlapping eligibility and rivalry.
\end{abstract}

%==============================================================================

\section{INTRODUCTION}
\vspace{-1mm}

Collective memory is shaped not only by what is recorded but by what remains visible. Platforms retain extensive records yet display only a fraction of them: search engines show top-ranked results, review sites highlight certain opinions, news outlets allocate limited airtime. Attention, not storage, is the scarce resource \cite{Simon1971,Falkinger2008}, so an event can lose influence while remaining on record. A restaurant facing a food-safety incident need not remove it; soliciting favourable reviews pushes it down the displayed list, and firms do pay to move visibility this way \cite{Mayzlin2014,LucaZervas2016}. The result is a contest: agents spend resources to preserve favourable events and suppress unfavourable ones, and since display capacity is fixed, what some events gain the rest lose.

We model this as the Memory Market Game (MMG). Each event carries an
objective importance and a reputational valence for the agents named in
it. Agents buy preservation and suppression effort under individual
budgets, one unit of suppression offsetting one unit of preservation, and
a softmax rule divides a fixed capacity across events. As the shares
have a common denominator, effort aimed at one event changes the
visibility of every other.

We make four contributions. First, mixed-strategy equilibria exist for the
general game and pure-strategy equilibria under quasi-concavity, while a
restricted two-agent contest admits neither an exact nor a weighted
potential. Second, with quadratic effort costs an explicit curvature
bound, independent of the number of events, gives equilibrium uniqueness
and geometric convergence of synchronous projected-gradient play. Third,
we characterize ranking reversals, the common rescaling of events that
receive no direct effort, and a sufficient memory-share guarantee for a
prescribed coordinated preservation policy. Fourth, over a finite design
grid we separate the mismatch of the no-effort allocation from the
displacement that strategic effort adds. The contest and coalition
results carry the restrictions stated with them; the design comparison
uses a numerical equilibrium-selection rule.

\textit{Related literature:}
Difference-form contest success functions map differences in effort into
relative success \cite{Hirshleifer1989,Skaperdas1996}; the softmax rule
used here shares the conditional-logit form \cite{McFadden1974} but
divides a fixed capacity rather than describing choice probabilities. Its
nearest cousin is the multi-battle contest, in which a budget is split
across fronts \cite{Roberson2006}; there fronts are won independently,
whereas one common denominator conserves capacity here. Reputation and
online-feedback research explains why visibility is worth paying for
\cite{Resnick2000,Dellarocas2003}.
A second stream studies exposure in rankings: fairness-of-exposure and
dynamic-ranking methods distribute attention by policy
\cite{SinghJoachims2018,Morik2020}, position auctions allocate slots by
score \cite{Edelman2007}, and feedback studies trace popularity-bias
amplification \cite{Mansoury2020,Klimashevskaia2024}. Nearest to us is
work in which the ranked items are themselves strategic
\cite{BenPoratTennenholtz2018,Hardt2016}, which asks how providers create
or shape new content under a given rule. We instead fix the record and the
rule and let budgeted agents move the scores of \emph{past} events in
either direction; because capacity is conserved, effort on one event
displaces events nobody targeted. Reader behaviour and the generation of
new events remain exogenous.
A third stream concerns equilibrium analysis: diagonal strict concavity
for uniqueness \cite{Rosen1965}, variational-inequality monotonicity for
contraction \cite{FacchineiPang2003}, and gradient play under
distributed, partial-information updates \cite{Tatarenko2021,Tan2024}.
Our curvature bound verifies those conditions for this game, at the cost
of certifying only a low-gain region.
\section{THE MEMORY MARKET GAME}
\label{sec:model}
\begin{figure*}[!t]
\centering
\resizebox{0.6
\textwidth}{!}{%
\begin{tikzpicture}[
  font=\sffamily\footnotesize, >=Latex,
  card/.style={draw=ink!65, rounded corners=3pt, thick, align=center, inner sep=3pt},
  chip/.style={card, minimum width=2.6cm, minimum height=0.45cm},
  flow/.style={-{Latex[length=2.2mm]}, line width=1.1pt, draw=ink!80},
  influence/.style={-{Latex[length=2.0mm]}, line width=1.0pt, dashed},
  heading/.style={font=\sffamily\bfseries\small, text=ink},
  note/.style={font=\scriptsize\itshape, text=ink!75, align=center}
]
% Stage headings
\node[heading] at (0,1.2)    {THE RECORD};
\node[heading] at (4.0,1.2)  {STRATEGIC ACTIONS};
\node[heading] at (8.3,1.2)  {LIMITED MEMORY};
\node[heading] at (12.6,1.2) {VISIBILITY OUTCOME};

% Events
\node[chip,fill=goldL,draw=goldA] (e1) at (0,0.6)  {\textbf{Positive review}\quad\textcolor{greenA}{+}};
\node[chip,fill=redL,draw=redA]   (e2) at (0,0)    {\textbf{Safety incident}\quad\textcolor{redA}{--}};
\node[chip,fill=blueL,draw=blueA] (e3) at (0,-0.6) {\textbf{Service history}\quad\textcolor{blueA}{+/--}};
\node[note] at (0,-1.1) {Events stay in the record};

% Agents
\node[card,fill=blueL,draw=blueA,minimum width=3.4cm,minimum height=0.75cm] (a1) at (4.0,0.42)
  {\textbf{Agent A}\\[-2pt]{\scriptsize\textcolor{blueA}{Preserve (+) / suppress (--)}}};
\node[card,fill=redL,draw=redA,minimum width=3.4cm,minimum height=0.75cm] (a2) at (4.0,-0.42)
  {\textbf{Agent B}\\[-2pt]{\scriptsize\textcolor{redA}{Preserve (+) / suppress (--)}}};
\node[note] at (3.5,-1.15) {Effort is costly\\and budget-limited};
\draw[flow,draw=blueA] (e1.east) -- (a1.west);
\draw[flow,draw=redA]  (e2.east) -- (a2.west);

% Platform with capacity bar
\node[card,fill=tealL,draw=tealA,minimum width=3.2cm,minimum height=1.65cm] (p) at (8.3,0) {};
\node[font=\small\bfseries,text=tealA] at (8.3,0.55) {THE PLATFORM};
\node[font=\scriptsize,text=ink] at (8.3,0.24) {Compares event scores};
\draw[rounded corners=2pt,fill=white,draw=tealA,line width=.8pt] (6.88,-0.2) rectangle (9.72,0.08);
\fill[goldA!85] (6.94,-0.15) rectangle (7.95,0.03);
\fill[redA!75]  (8.00,-0.15) rectangle (8.70,0.03);
\fill[blueA!75] (8.75,-0.15) rectangle (9.66,0.03);
\node[font=\scriptsize,text=ink,align=center] at (8.3,-0.55) {Fixed capacity: more for one\\event means less for others};
\draw[influence,draw=blueA] (a1.east) -- (a1.east -| p.west);
\draw[influence,draw=redA]  (a2.east) -- (a2.east -| p.west);

% Outcomes
\node[chip,fill=greenL,draw=greenA,minimum width=2.7cm] (v1) at (12.6,0.6)  {\textbf{Most visible}\quad\textcolor{greenA}{1}};
\node[chip,fill=goldL,draw=goldA,minimum width=2.7cm]   (v2) at (12.6,0)    {\textbf{Less visible}\quad 2};
\node[chip,fill=redL,draw=redA,minimum width=2.7cm]     (v3) at (12.6,-0.6) {\textbf{Fades from view}\quad 3};
\node[note] at (12.6,-1.1) {Visibility shapes reputation};
\draw[flow,draw=tealA] (p.east) -- (v2.west);

% Feedback loop
\draw[influence,draw=blueA,line width=1.1pt]
  (12.6,-1.3) -- (12.6,-1.55) -| (6.1,-1.55) -- (5.35,-0.8);
\node[fill=white,inner sep=2pt,font=\scriptsize\bfseries,text=blueA] at (9.3,-1.55)
  {Visibility determines agents' reputational payoffs};
\end{tikzpicture}%
}
\vspace{-1mm}
\caption{The Memory Market Game}
\label{fig:model}
\end{figure*}
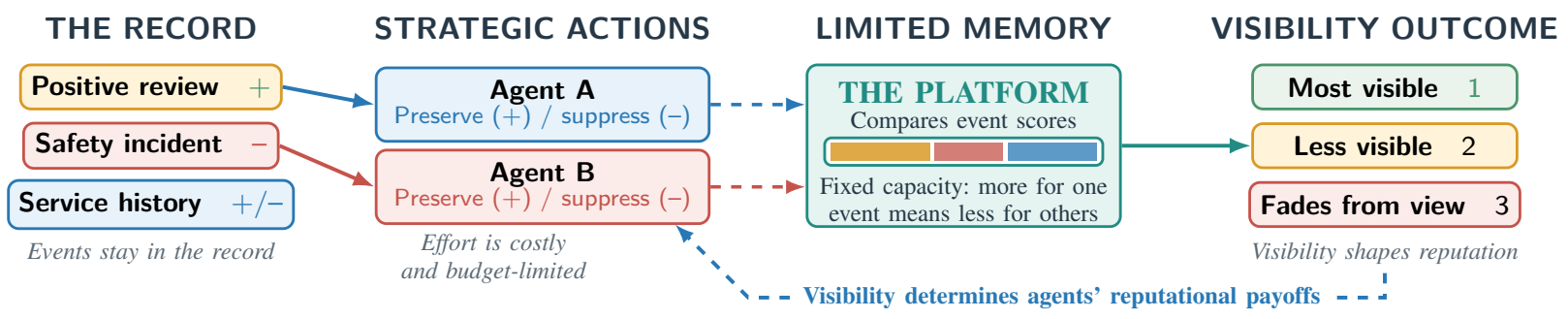
\vspace{-2mm}
\subsection{Agents, events and capacity}
\vspace{-1mm}
Let $\NN=\{1,\dots,n\}$ be a finite set of self-interested agents and
$\E=\{e_1,\dots,e_m\}$, $m\ge2$, a finite set of past events. Agent $i$ acts
on events in $\E_i\subseteq\E$, with effort outside $\E_i$ fixed at zero.
Each event $e$ carries two exogenous attributes, fixed when it occurs: an
\emph{objective importance} $\omega_e\in\R_+:=[0,\infty)$, so zero importance
is permitted, and a \emph{reputational valence} $\theta_e\in\R$, positive for
events that reflect well on the agents named in them. We write $i\in e$ when
agent $i$ is named in event $e$, an exogenous relation. The system has a
bounded \emph{memory capacity} $M>0$ to distribute over $\E$, a divisible
attention budget, so $x_e/M$ is the share assigned to $e$. The model captures
visibility, not deletion: an event may remain in the record with a small
share. Figure~\ref{fig:model} summarises the decision sequence and the
feedback from visibility to incentives.

\emph{Memory scores:} Agent $i$ chooses preservation and suppression
efforts $r_i(e),f_i(e)\ge0$ for $e\in\E_i$, and zero effort on all other
events. Write $\sigma=(r_i,f_i)_{i\in\NN}$; each event $e$ acquires a
score
\begin{align}\label{eq:score}
S_e(\sigma)\;=\;\omega_e+\sum_{i\in\NN}r_i(e)-\sum_{i\in\NN}f_i(e),
\end{align}
the sum of its importance and net effort. Define
$\nu_e:=\sum_i r_i(e)-\sum_i f_i(e)$, so $S_e=\omega_e+\nu_e$. Only net
effort matters: one unit of suppression offsets one unit of preservation.
This unit-for-unit efficacy is a normalization, and an asymmetric score
$S_e=\omega_e+\sum_i r_i(e)-\kappa\sum_i f_i(e)$, $\kappa>0$, is outside the
present model.

\emph{Market clearing:} Once the scores are set, the capacity $M$ is divided
among the events by a softmax rule with gain $\lambda>0$:
\begin{align}\label{eq:clear}
x_e(\sigma)\;=\;M\,\frac{\exp\!\big(\lambda S_e(\sigma)\big)}
{\sum_{k\in\E}\exp\!\big(\lambda S_k(\sigma)\big)}.
\end{align}
Rule~\eqref{eq:clear} is a difference-form contest success function
\cite{Hirshleifer1989} and a logit share model \cite{McFadden1974}, with two
properties used throughout: \emph{capacity is conserved}, so memory given to
one event is taken from the others, which separates the memory market from a
contest over independent prizes; and \emph{only score differences matter}, so
a lead in importance is reversed by an equal amount of net effort, with
$\lambda>0$ setting how sharply shares respond.

\subsection{Budgets, payoffs and equilibrium}
\vspace{-1mm}
\emph{Budgets:} Effort is measured in normalized units and bought at a convex
cost with exponent $\tau\ge0$; comparisons across $\tau$ hold that unit and
the coefficients $c_r,c_f$ fixed. Agent $i$'s total spend and feasible set:
\begin{align}\label{eq:budget}
T^\tau_i(\sigma)&=\sum_{e\in\E_i}\!\big(c_r r_i(e)^{1+\tau}+c_f f_i(e)^{1+\tau}\big),\\
\Sigma_i^\tau&=\{\sigma_i:T_i^\tau(\sigma_i)\le B_i\},\nonumber
\end{align}
with unit costs $c_r,c_f>0$ and budget $B_i<\infty$: $B_i$ bounds agent $i$'s
endowment, while $T_i^\tau$ is the opportunity cost of deploying it and is
also deducted from utility. At the baseline $\tau=0$, \eqref{eq:budget} is the
linear constraint $c_r\|r_i\|_1+c_f\|f_i\|_1\le B_i$; $\tau>0$ changes the
cost curvature considered in Section~\ref{sec:mech}. Write
$\Sigma^\tau=\prod_i\Sigma_i^\tau$.

\emph{Payoffs and stakes:} Agent $i$ values the visibility of event $e$
according to its marginal stake $\gamma_i(e)$. Its payoff is
\begin{equation}\label{eq:util}
U_i(\sigma)=\sum_{e\in\E}\gamma_i(e)x_e(\sigma)-T_i^\tau(\sigma).
\end{equation}
For a reputational interpretation, let $\rho:=\alpha/\delta>0$ be the
ratio of the reputation-value coefficient to the reputation-decay rate.
Then
\begin{equation}\label{eq:stake}
\gamma_i(e):=\rho\theta_e\left[
\mathbf 1_{\{i\in e\}}-
\sum_{k\ne i}\beta_{ik}\mathbf 1_{\{k\in e\}}
\right].
\end{equation}
Only the ratio $\rho$ enters the static game. The coefficients
$\beta_{ik}\ge0$ measure how much agent $i$'s payoff falls when a rival $k$
is associated with a visible event; note that $\rho$ multiplies the whole
bracket in \eqref{eq:stake}, so a rival term contributes $\rho\beta_{ik}$, not
$\beta_{ik}$. Write $g_i(e):=|\gamma_i(e)|$ for the \emph{stake} and
$s_i(e):=\operatorname{sign}\gamma_i(e)$; a negative stake means $i$ gains
from \emph{lower} visibility of $e$, and $\gamma_i(e)=0$ means indifference.
Agents $i$ and $j$ have \emph{opposed} stakes in $e$ when
$s_i(e)s_j(e)=-1$. Results involving a stake assume $g_i(e)>0$; additive
payoff constants are omitted, as they do not affect best responses.

\begin{definition}[MME]\label{def:mme}
A profile $\sigma^\star\in\Sigma^\tau$ is a \emph{pure-strategy Memory
Market Equilibrium} (MME) if
$U_i(\sigma^\star_i,\sigma^\star_{-i})\ge
U_i(\sigma_i,\sigma^\star_{-i})$ for every $i\in\NN$ and every
$\sigma_i\in\Sigma_i^\tau$; mixed-strategy equilibrium is defined as usual
on distributions over the feasible sets.
\end{definition}

Throughout we assume \textbf{(A1)} $0<B_i<\infty$ for every $i$ and
\textbf{(A2)} $c_r,c_f>0$, $\lambda>0$; the general results allow any
eligibility sets $\E_i$, and \textbf{(A3)}--\textbf{(A5)} are stated where
used. At a best response $\min\{r_i(e),f_i(e)\}=0$, since reducing both
efforts by their minimum leaves the scores unchanged and lowers cost.

%==============================================================================
\section{STRUCTURE AND DYNAMICS}
\label{sec:eq}
The clearing rule determines how changes in event scores affect their visibility. Since the platform has a fixed capacity, an increase in one event's allocation reduces the allocation available to the others. Lemma \ref{lem:clear} establishes these properties and derives the marginal effects of changes in event scores.
\begin{lemma}[Clearing rule]\label{lem:clear}
For every $\sigma$ and every $e\in\E$ the rule \eqref{eq:clear} satisfies
(i) $0<x_e<M$ and $\sum_{k}x_k=M$ exactly, so the capacity always binds;
(ii) $x$ is real analytic in $\sigma$;
(iii) $\partial x_e/\partial S_e=\lambda x_e(1-x_e/M)>0$ and
$\partial x_e/\partial S_k=-\lambda x_ex_k/M<0$ for $k\ne e$;
(iv) $0<\partial x_e/\partial S_e\le\lambda M/4$, with equality iff
$x_e=M/2$;
(v) $|\partial^2 x_e/\partial S_p\partial S_q|\le2\lambda^2x_e$ for all
$p,q$; and
(vi) for every $p$, $\sum_{q\in\E}|\partial^2x_e/\partial S_p\partial S_q|
\le\frac{5}{2}\lambda^2x_e$.
\end{lemma}

\begin{proof}
Parts (i)--(iv) follow from \eqref{eq:clear} and the quotient rule, with
$0<x_e<M$ since $m\ge2$, and $\lambda x_e(1-x_e/M)$ having maximum
$\lambda M/4$ at $x_e=M/2$. For (v), put $\rho_k=x_k/M$ and differentiate
twice:
\[
\frac{\partial^2x_e}{\partial S_p\partial S_q}
=M\lambda^2\rho_e\Big[(\delta_{ep}-\rho_p)(\delta_{eq}-\rho_q)
-\delta_{pq}\rho_p+\rho_p\rho_q\Big].
\]
Since $0<\rho_k<1$, the first product in the bracket has absolute value
at most one, and the remaining term
$-\delta_{pq}\rho_p+\rho_p\rho_q$ also has absolute value at most one.
Thus the bracket has absolute value at most two; using
$M\rho_e=x_e$ gives (v). For (vi), fix $p$ and sum over $q$: using
$\sum_q|\delta_{eq}-\rho_q|=2(1-\rho_e)\le2$ and
$\sum_q|\rho_p\rho_q-\delta_{pq}\rho_p|=2\rho_p(1-\rho_p)\le\frac12$,
\[
\sum_{q\in\E}\Big|\frac{\partial^2x_e}{\partial S_p\partial S_q}\Big|
\le M\lambda^2\rho_e\Big[2|\delta_{ep}-\rho_p|+\tfrac12\Big]
\le\frac{5}{2}\lambda^2x_e . \qedhere
\]
\end{proof}

Part (vi) is the bound that matters for Theorem~\ref{thm:mono}: proportional
to $x_e$, it costs a factor $\sum_ex_e=M$ when summed over events, not $mM$.

\begin{proposition}[No potential]\label{prop:nopot}
Let $i\ne j$ have $\gamma_i(e)=g_i>0$, $\gamma_j(e)=-g_j<0$ and
$\gamma_i(k)=\gamma_j(k)=0$ for $k\ne e$, and fix all effort coordinates
except $a=r_i(e)$ and $b=f_j(e)$ on $R=[0,\bar a]\times[0,\bar b]$. The
resulting two-agent game admits neither an exact nor a weighted potential
with positive weights \cite{MondererShapley1996}.
\end{proposition}

\begin{proof}
With the other scores fixed, write $x_e(z)=M e^{\lambda(\omega_e+z)}/(e^{\lambda(\omega_e+z)}+K)$ for $K>0$. Up to terms free of $(a,b)$, $U_i=g_ix_e(a-b)-c_ra$ and
$U_j=-g_jx_e(a-b)-c_fb$. An exact potential $P$ would force
$(g_i+g_j)\iint_Rx_e''(a-b)\,db\,da=0$ for every subrectangle. From
\eqref{eq:loc}, $x_e''(z)=\lambda^2x_e(1-x_e/M)(1-2x_e/M)$ vanishes only
at $x_e=M/2$, hence at a single $z$ as $x_e$ is strictly increasing, so a
subrectangle on which $x_e''$ has a strict sign gives a contradiction. For
weights $w_i,w_j>0$, equality of the mixed second derivatives of $P$
requires $-g_i/w_i=g_j/w_j$ wherever $x_e''\ne0$, impossible for positive
$g$'s and $w$'s.
\end{proof}
Any potential for the full game would restrict to a potential of the same type in the restricted game. Thus, under the stated hypotheses, exact and positive-weight weighted potentials are ruled out. Ordinal potentials are not excluded. We therefore study equilibrium using monotonicity rather than potential methods.

\begin{theorem}[Existence]\label{thm:exist}
Let \textbf{(A1)}--\textbf{(A2)} hold and $\tau\ge0$. Then a mixed-strategy
Nash equilibrium exists. If, in addition, every $U_i(\cdot,\sigma_{-i})$ is
quasi-concave on $\Sigma_i^\tau$, a pure-strategy MME exists.
\end{theorem}

\begin{proof}
Each $T_i^\tau$ is continuous and convex on $\mathbb R_+^{2m}$, so
$\Sigma_i^\tau$, and hence $\Sigma^\tau$, is nonempty, compact and convex,
while each $U_i$ is continuous on $\Sigma^\tau$ by
Lemma~\ref{lem:clear}(ii). Glicksberg's theorem \cite{Glicksberg1952}
gives a mixed-strategy equilibrium. Under quasi-concavity each
$BR_i(\sigma_{-i})$ is convex, and nonempty, compact-valued and upper
hemicontinuous by compactness of $\Sigma_i^\tau$ and continuity of $U_i$,
so Kakutani's theorem applied to $\prod_i BR_i$ gives a pure-strategy MME;
see also \cite{Rosen1965}.
\end{proof}

Quasi-concavity is not a mild addition: it can fail outright, as the position
of the feasible interval shows.

\begin{remark}[Own-payoffs can be S-shaped]\label{rem:sshape}
Let $\tau=0$, vary one own-effort coordinate $y$ over $[0,\bar y]$ under
\textbf{(A3)}--\textbf{(A4)}, and put
$\psi(y):=x_e(y)\big(1-x_e(y)/M\big)$. The restricted payoff $g_ix_e(y)-cy$
has derivative $g_i\lambda\psi(y)-c$ by Lemma~\ref{lem:clear}(iii), and
$\psi$ is strictly unimodal with maximum $M/4$ where $x_e=M/2$. If that point
is interior to $[0,\bar y]$ and
\begin{equation}\label{eq:sshape}
g_i\lambda M/4\;>\;c\;>\;g_i\lambda\max\{\psi(0),\psi(\bar y)\},
\end{equation}
then the derivative is negative, then positive, then negative: the payoff is
strictly S-shaped and not quasi-concave, so Theorem~\ref{thm:exist} gives
only a mixed-strategy equilibrium. Otherwise it changes sign at most once and
quasi-concavity may hold; the baseline model does not guarantee it.
\end{remark}
Quasi-concavity therefore cannot be assumed at $\tau=0$;
Lemma~\ref{lem:corner} instead gives corner conditions, and strict cost
convexity restores concavity, which Theorem~\ref{thm:mono} exploits at
$\tau=1$. Between those conditions the best response is interior and is not
characterized here; Section~\ref{sec:num} reports a design point there.

\subsection{Interior and corner conditions in a contest}
\vspace{-1mm}
Fix an event $e$ and agents $i\ne j$ with opposed stakes $g_i,g_j>0$ in $e$.
Two restrictions apply below.

\medskip\noindent\textbf{(A3) Single-event stake.} Neither $i$ nor $j$ holds
a stake in any event other than $e$, and every other effort coordinate of
$\sigma$ is held fixed.

\noindent\textbf{(A4) Scalar control.} Each of $i$ and $j$ chooses one
scalar effort, with every other coordinate of its own strategy fixed at
zero, so that its feasible set is an interval.
\medskip

\noindent Under \textbf{(A3)} the marginal payoff from a change in $S_e$
involves only $x_e$; \textbf{(A4)} makes each feasible set an interval. Write
$a:=r_i(e)$, $b:=f_j(e)$, $z:=a-b$, and let $K>0$ be the combined score
exponential of the other events; \eqref{eq:clear} restricts to the
logistic
\begin{equation}\label{eq:loc}
x_e(z)=M\,\frac{e^{\lambda(\omega_e+z)}}{e^{\lambda(\omega_e+z)}+K},
\end{equation}
and at $\tau=0$ the restricted payoffs are $\varphi_i(a,b)=g_ix_e(z)-c_ra$
and $\varphi_j(a,b)=-g_jx_e(z)-c_fb$ on $[0,\bar a]\times[0,\bar b]$,
$\bar a=B_i/c_r$, $\bar b=B_j/c_f$. We call this a \emph{contest} over $e$.

\begin{lemma}[Interior necessity, corner sufficiency]\label{lem:corner}
At $\tau=0$, under \textbf{(A2)}, \textbf{(A3)} and \textbf{(A4)}, in any contest:
\begin{enumerate}\itemsep0.5pt
\item[(i)] a jointly interior pair $(a,b)\in(0,\bar a)\times(0,\bar b)$ can be
mutually optimal only if $g_i/c_r=g_j/c_f$; for fixed costs this is a line
through the origin in $\R_+^2$, of Lebesgue measure zero;
\item[(ii)] if $g_i\lambda M/4<c_r$ then $a^\star=0$ for every $b$, so $i$'s
effort is dominated; the analogue holds for $j$ with $(g_j,c_f)$;
\item[(iii)] if $g_i\lambda\underline\psi>c_r$, where
$\underline\psi:=\min_{z\in[-\bar b,\bar a]}x_e(z)\big(1-x_e(z)/M\big)$, then
$a^\star=\bar a$ for every $b$; the analogue holds for $j$.
\end{enumerate}
\end{lemma}

\begin{proof}
From \eqref{eq:loc}, $x_e'(z)=\lambda x_e(z)(1-x_e(z)/M)>0$, so
$\partial\varphi_i/\partial a=g_ix_e'(z)-c_r$ and
$\partial\varphi_j/\partial b=g_jx_e'(z)-c_f$. (i) If both efforts are
interior at a mutual best response, the first-order conditions give
$g_ix_e'(z)=c_r$ and $g_jx_e'(z)=c_f$, hence $g_i/c_r=g_j/c_f$; for fixed
$c_r,c_f$ this defines a line in the $(g_i,g_j)$-plane, of Lebesgue
measure zero. (ii) By Lemma~\ref{lem:clear}(iv),
$g_ix_e'(z)\le g_i\lambda M/4<c_r$ for every feasible $z$, so
$\partial\varphi_i/\partial a<0$ throughout $[0,\bar a]$ and $a=0$ is the
unique best response. (iii) By definition of $\underline\psi$,
$\partial\varphi_i/\partial a\ge g_i\lambda\underline\psi-c_r>0$, so
$\varphi_i$ is strictly increasing and $a=\bar a$ is the unique best
response. The arguments for $j$ are identical.
\end{proof}
Part (i) is a necessary condition on parameters, not a sufficient one: on the
line $g_i/c_r=g_j/c_f$ neither a jointly interior equilibrium nor one with a
single interior agent is excluded. Neither is characterized here, and nothing
above shows that interior equilibria are impossible.

\begin{corollary}[Pure equilibrium at strict best-response corners]
\label{cor:corner}
Suppose each agent $i$ has one active beneficial scalar effort
$y_i\in[0,B_i/c_i]$ on an event $e(i)$ and has no stake in other
events. Let
\[
\underline\psi_i:=
\min_{\sigma\in\Sigma^0}
x_{e(i)}(\sigma)\bigl(1-x_{e(i)}(\sigma)/M\bigr),
\]
where $\Sigma^0$ is the joint feasible set at $\tau=0$. If, for every
agent, either $g_i\lambda M/4<c_i$ or
$g_i\lambda\underline\psi_i>c_i$, then the profile assigning zero
effort in the first case and full effort in the second is a pure MME.
\end{corollary}

\begin{proof}
For every feasible profile of the other agents,
the derivative of agent $i$'s payoff with respect to its beneficial
effort is $g_i\lambda x_{e(i)}(1-x_{e(i)}/M)-c_i$.
The first condition makes this derivative strictly negative throughout
the feasible set; the second makes it strictly positive. Each prescribed
corner is therefore the unique best response for every opponents' profile,
so the resulting profile is a pure MME.
\end{proof}

\subsection{Projected gradient play}
\vspace{-1mm}
As a benchmark, if each agent controls one scalar $y_i\in[0,\bar y_i]$ with
$\partial U_i/\partial y_i\ge m_i>0$ throughout, projected gradient play
\cite{Tatarenko2021,Tan2024} drives every coordinate to its cap in at most
$\lceil\bar y_i/(\eta m_i)\rceil$ steps, resolving no strategic interaction.
The next result is the paper's main dynamic statement, for the quadratic-cost
game under synchronous, exact, full-information updates; delayed,
asynchronous or inexact play is not covered.

\begin{theorem}[Unique MME and global convergence]\label{thm:mono}
Let $\tau=1$, put $c_{\min}:=\min\{c_r,c_f\}$,
$c_{\max}:=\max\{c_r,c_f\}$, $\hat g:=\max_{i,e}g_i(e)$, and
\begin{equation}\label{eq:Lconst}
L:=5n\lambda^2M\hat g .
\end{equation}
If $2c_{\min}>L$, then the pseudo-gradient
$F(\sigma):=(\partial U_i/\partial\sigma_i)_{i\in\NN}$ satisfies
\(
\langle F(\sigma)-F(\sigma'),\sigma-\sigma'\rangle
\le -\mu\|\sigma-\sigma'\|^2,
\)
where $\mu:=2c_{\min}-L>0$. The MMG then has a unique MME $\sigma^\star$,
and projected-gradient play
$\sigma^{t+1}=\operatorname{Proj}_{\Sigma^1}\{\sigma^t+\eta F(\sigma^t)\}$
with any constant step $0<\eta<2\mu/\ell^2$, where
$\ell:=L+2c_{\max}$, satisfies
\begin{equation}\label{eq:rate}
\|\sigma^{t}-\sigma^\star\|\le q^{\,t}\|\sigma^{0}-\sigma^\star\|,
\qquad q:=\sqrt{1-2\eta\mu+\eta^2\ell^2}<1,
\end{equation}
for every $\sigma^{0}\in\Sigma^1$.
\end{theorem}

\begin{proof}
Index the $d$ effort coordinates by $p,q$; let $e(p)$ be the event
affected by coordinate $p$, $i(p)$ the agent choosing it, and
$\zeta_p=\pm1$ according as $p$ preserves or suppresses. The memory part
of the $p$th pseudo-gradient component and its derivative in $q$ are

\vspace{-3mm}
{\small\[
F_p^{\mathrm{mem}}
=\zeta_p\!\sum_{e\in\E}\!\gamma_{i(p)}(e)
\frac{\partial x_e}{\partial S_{e(p)}},\
J^{\mathrm{mem}}_{pq}
=\zeta_p\zeta_q\!\sum_{e\in\E}\!\gamma_{i(p)}(e)
\frac{\partial^2x_e}{\partial S_{e(p)}\partial S_{e(q)}}.
\]}
The signs enter only through $\zeta_p\zeta_q$ and do not affect an
absolute-value bound. We bound the induced $\infty$- and $1$-norms, that
is the maximum row sum and the maximum column sum, rather than the
Frobenius norm; this is what keeps the constant free of $m$ and $d$.
Each event is affected by at most two effort coordinates per agent, one
for preservation and one for suppression, so at most $2n$ coordinates
have $e(p)$, or $e(q)$, equal to any given event.

\emph{Row sums.} Fix a row $p$ and partition the columns $q$ by the event
$e(q)$ they act on; each class has at most $2n$ members and shares the
same inner second-derivative index. Hence
\begin{align*}
\sum_{q=1}^{d}|J^{\mathrm{mem}}_{pq}|
&\le 2n\sum_{k\in\E}\sum_{e\in\E} g_{i(p)}(e)
 \Big|\frac{\partial^2x_e}{\partial S_{e(p)}\partial S_{k}}\Big|\\
&\le 2n\sum_{e\in\E}g_{i(p)}(e)\,\tfrac52\lambda^2x_e
 \le 5n\lambda^2\hat g\sum_{e\in\E}x_e=L,
\end{align*}
where the second inequality is Lemma~\ref{lem:clear}(vi) applied with
$p$-index $e(p)$ and summation over the column event $k$, and the last
step uses $\sum_ex_e=M$ from Lemma~\ref{lem:clear}(i).

\emph{Column sums.} Fix a column $q$ and partition the rows $p$ by
$e(p)$, again at most $2n$ per class. The softmax Hessian is symmetric in
its two score indices, so with the column event $e(q)$ fixed the same
application of Lemma~\ref{lem:clear}(vi), now summing over the row event,
gives $\sum_{p=1}^{d}|J^{\mathrm{mem}}_{pq}|\le L$ by the identical
computation with $|\gamma_{i(p)}(e)|\le\hat g$.

Hence $\|J^{\mathrm{mem}}\|_1\le L$ and
$\|J^{\mathrm{mem}}\|_\infty\le L$, so
$\|J^{\mathrm{mem}}\|_2\le
(\|J^{\mathrm{mem}}\|_1\|J^{\mathrm{mem}}\|_\infty)^{1/2}\le L$.
The Jacobian need not be symmetric because agents can have different
stakes. The spectral-norm bound nevertheless gives
$|v^\top J^{\mathrm{mem}}(\sigma)v|\le L\|v\|^2$ for every $v$.

The quadratic cost contributes a diagonal Hessian whose entries are either
$-2c_r$ or $-2c_f$, so
$v^\top J(\sigma)v\le|v^\top J^{\mathrm{mem}}(\sigma)v|-2c_{\min}\|v\|^2\le
-\mu\|v\|^2$ for every $v$. Integrating this inequality along the
segment joining $\sigma$ and $\sigma'$ gives strong anti-monotonicity of
$F$, and restricting it to one agent's block gives strict concavity of
each payoff in its own strategy.

For uniqueness, $\Sigma^1=\prod_i\Sigma^1_i$ is a product of the
individual quadratic budget sets, each nonempty, compact and convex, so
the MMG is a concave game on a convex product set and its equilibria are
exactly the solutions of the variational inequality
$\langle-F(\sigma^\star),\sigma-\sigma^\star\rangle\ge0$ for all
$\sigma\in\Sigma^1$. Strong anti-monotonicity makes $-F$ strongly
monotone, so this variational inequality has exactly one solution
\cite{FacchineiPang2003}; equivalently, the inequality implies Rosen's
diagonal strict concavity \cite{Rosen1965}. Compactness and concavity give
existence, so the MME exists and is unique.

For the rate, the same bound gives $\|J\|_2\le L+2c_{\max}=\ell$, so $F$
is $\ell$-Lipschitz. The unprojected map satisfies
$\|\sigma-\sigma'+\eta(F(\sigma)-F(\sigma'))\|^2\le
(1-2\eta\mu+\eta^2\ell^2)\|\sigma-\sigma'\|^2$, whose factor is below one
exactly for $0<\eta<2\mu/\ell^2$. Projection onto $\Sigma^1$ is
nonexpansive, so the projected map is a contraction of modulus $q$ in
norm with unique fixed point $\sigma^\star$, which gives \eqref{eq:rate}.
\end{proof}
\begin{table}[!t]
\centering\footnotesize
\setlength{\tabcolsep}{4pt}
\caption{Best responses in the strategic-amnesia example.}
\label{tab:amnesia-br}
\vspace{-1mm}
\begin{tabular}{lccc}
\toprule
Agent & Beneficial actions & Budget & Best-response effort\\
\midrule
$A$ & suppress $e$ / preserve $f$ & $8$ & $8$\\
$B$ & preserve $f$ / suppress $e$ & $2$ & $2$\\
$C$ & preserve $f$ / suppress $e$ & $1$ & $1$\\
\bottomrule
\end{tabular}
\end{table}
Theorem~\ref{thm:mono} gives a global contraction condition with an explicit
step-size range and geometric rate; $q$ in~\eqref{eq:rate} is the modulus in
norm, so squared error contracts by $q^2$ per step. The constant $L$ is
independent of the number of events, because the Hessian bound is
proportional to $x_e$ and $\sum_e x_e=M$, and grows linearly in the number of
agents $n$ and the largest stake $\hat g$ and quadratically in the clearing
gain $\lambda$. Simultaneous preservation and suppression are dominated at a
best response, but projected-gradient iterates need not respect that, so the
bound counts both coordinates. With $M=100$, $\hat g=5$ and unit costs the
condition becomes $\lambda\le0.0163$: it certifies the low-gain region, not
the baseline $\lambda=0.5$, where the iteration may still converge without a
global guarantee. Section~\ref{sec:num} also tests an
overlapping-eligibility instance inside the certified region.

%==============================================================================
\section{DISTORTIONS AT EQUILIBRIUM}
\label{sec:distort}
\vspace{-1mm}
\subsection{Strategic amnesia}
\vspace{-1mm}
An event of higher objective importance need not receive a larger share at
equilibrium, as the following three-agent, two-event instance shows.

\begin{theorem}[Strategic amnesia]\label{thm:amnesia}
There exists an MMG instance satisfying \textbf{(A1)}--\textbf{(A2)} with
$\tau=0$, an MME $\sigma^\star$, and two events $e,f$ with $\omega_e>\omega_f$
yet $x_e(\sigma^\star)<x_f(\sigma^\star)$.
\end{theorem}

\begin{proof}
Take $\NN=\{A,B,C\}$, $\E=\{e,f\}$, $M=100$, $\lambda=1/2$, $\alpha=1$ and
$\delta=1/5$, so $\rho=5$. Agent $A$ is the only agent named in $e$, which is major and
damaging to it ($\omega_e=10$, $\theta_e=-1$); $B$ and $C$ are named in
$f$, which is minor and favourable to them ($\omega_f=8$, $\theta_f=+1$);
every $\beta_{ik}=0$. By \eqref{eq:stake}, agent $A$ has a negative stake of magnitude $5$
in $e$, so it gains from lowering the visibility of $e$, while $B$ and $C$
have positive stakes of magnitude $5$ in $f$.
All other stakes are zero. Each agent can act on both events:
$\E_A=\E_B=\E_C=\{e,f\}$. Since there are only two events and
$c_r=c_f=1$, each agent can improve its payoff either by suppressing the
event it opposes or by preserving the other. The two are equivalent here only
because, by \eqref{eq:score}, each moves $S_e-S_f$ by the same amount under
the unit-for-unit normalization and, with $c_r=c_f$ and $\tau=0$, costs the
same; with $c_r\ne c_f$, $\tau>0$, or more than two events they are not
interchangeable. The opposite actions reduce payoffs, so each agent's problem
is scalar: $y_A=f_A(e)+r_A(f)$ and $y_i=r_i(f)+f_i(e)$ for $i=B,C$, each
ranging over that agent's budget, with the logistic form \eqref{eq:loc}.
Table~\ref{tab:amnesia-br} summarizes the equivalent beneficial actions
and their best responses.

Set $B_A=8$, $B_B=2$ and $B_C=1$, and consider the profile with
$f_A(e)=8$, $r_B(f)=2$, $r_C(f)=1$ and all other efforts zero. Then
$S_e=2$, $S_f=11$, and \eqref{eq:clear} gives
$x_e=1.0987$ and $x_f=98.9013$. To verify global optimality, vary each
agent's total beneficial effort over its full budget. For $A$, $x_e$
ranges from $1.0987$ to $37.754$, and for $B$ and $C$, $x_f$ remains
above $M/2$ throughout their feasible ranges. In each case the minimum
of $x(1-x/M)$ occurs at the full-budget endpoint and equals
$\underline\psi=1.0866$. Hence
$g\lambda\underline\psi=2.7166>1$, so Lemma~\ref{lem:corner}(iii)
shows that spending the full budget is optimal. The allocation is
therefore supported by global best responses. The best response need
not be unique: any split of the stated total effort between the two
beneficial coordinates has the same effect on the score difference and
the same cost.
\end{proof}

Without effort the two events receive $73.11$ and $26.89$ units, against
$1.10$ and $98.90$ at $\sigma^\star$; preservation of $f$ alone ($37.75$,
$62.25$) and suppression of $e$ alone ($4.74$, $95.26$) each reverse the
ordering by themselves. This is an instance of an exact criterion: write
$\nu_e(\sigma):=\sum_ir_i(e)-\sum_if_i(e)$, so $S_e=\omega_e+\nu_e$.

\begin{corollary}[When ranking reverses]\label{cor:rev}
For any profile $\sigma$ and any two events $e,f$ with
$\omega_e>\omega_f$,
\begin{equation}\label{eq:revcrit}
x_e(\sigma)<x_f(\sigma)\iff \nu_f(\sigma)-\nu_e(\sigma)>\omega_e-\omega_f ,
\end{equation}
a threshold involving neither $\lambda$ nor $M$. Suppose in addition that
$\tau=0$, that every agent with a stake in $e$ or $f$ has that event as
its only active coordinate, in the sense of \textbf{(A3)}--\textbf{(A4)},
that it satisfies the strict condition of Lemma~\ref{lem:corner}(iii)
there, and that

\medskip\noindent\textbf{(A5) Aligned stakes.} every such agent's
beneficial action is preservation of $f$ or suppression of $e$;
equivalently, no agent holds a positive stake in $e$ or a negative stake
in $f$, so no active coordinate raises $\nu_e$ or lowers $\nu_f$.

\medskip\noindent Then each such agent spends its whole budget on that
coordinate and
\begin{equation}\label{eq:nugap}
\nu_f-\nu_e=\sum_{i\in P_f}B_i/c_r+\sum_{i\in S_e}B_i/c_f ,
\end{equation}
with $P_f$ preserving $f$ and $S_e$ suppressing $e$: an explicit condition
on budgets, costs and importance.
\end{corollary}

\begin{proof}
Since $t\mapsto e^{\lambda t}$ is increasing and the denominator in
\eqref{eq:clear} is common, $x_e<x_f$ iff $S_e<S_f$, i.e.
$\omega_e+\nu_e<\omega_f+\nu_f$; rearranging removes $\lambda$ and $M$ and
gives \eqref{eq:revcrit}.
Under the stated restriction each agent has a single feasible coordinate
and Lemma~\ref{lem:corner}(iii) puts it at its cap. By \textbf{(A5)} every
such coordinate contributes with a positive sign to $\nu_f-\nu_e$ and none
contributes negatively, so the signed sum reduces to \eqref{eq:nugap}.
\end{proof}

\noindent Without \textbf{(A5)} the right-hand side of \eqref{eq:nugap} is a
signed sum: an agent with a positive stake in $e$, or a negative stake in
$f$, subtracts its capped effort. The criterion \eqref{eq:revcrit} is
unconditional. In the example $\omega_e-\omega_f=2$ and $\nu_f-\nu_e=11$, so
preservation of $f$ and suppression of $e$ reverse the visibility ranking;
the budget interpretation uses (A3)--(A5).

\subsection{Allocation changes for bystander events}
\vspace{-1mm}
For a bystander event that receive no direct effort, their scores remain unchanged, and their visibility changes only through adjustments to the common denominator of the softmax allocation.

\begin{proposition}[Common crowding factor]\label{prop:crowd}
Let $\mathcal{U}\subseteq\E$ be the set of \emph{bystander} events on which no
agent spends effort, so $S_k(\sigma)=\omega_k$ $\forall k\in\mathcal{U}$,
let $\mathcal{C}:=\E\setminus\mathcal{U}$ be the set of \emph{contested}
events, and let $D(\sigma):=\sum_{j\in\E}e^{\lambda S_j(\sigma)}$. Then
\begin{enumerate}\itemsep0.5pt
\item[(i)] every bystander is rescaled by one common factor,
\begin{equation}\label{eq:kappa}
\frac{x_k(\sigma)}{x_k(\mathbf{0})}=\kappa(\sigma):=
\frac{D(\mathbf{0})}{D(\sigma)}\qquad\text{for all }k\in\mathcal{U},
\end{equation}
\item[(ii)] relative standing among bystanders is invariant to all strategic
effort: $x_k(\sigma)/x_l(\sigma)=e^{\lambda(\omega_k-\omega_l)}$ 
$\forall k,l\in\mathcal{U}$;
\item[(iii)] memory is conserved, so what the bystanders lose is exactly what
the contested events gain,
\begin{equation}\label{eq:conserve}
\sum_{k\in\mathcal{U}}\!\big[x_k(\mathbf{0})-x_k(\sigma)\big]
=\sum_{e\in\mathcal{C}}\!\big[x_e(\sigma)-x_e(\mathbf{0})\big].
\end{equation}
\end{enumerate}
\end{proposition}

\begin{proof}
For $k\in\mathcal{U}$ the numerator of \eqref{eq:clear} is
$e^{\lambda\omega_k}$ at every profile, so
$x_k(\sigma)=Me^{\lambda\omega_k}/D(\sigma)$; dividing by
$x_k(\mathbf{0})$ gives \eqref{eq:kappa} with the same factor for every
$k$, which is (i). The quotient of $x_k$ and $x_l$ cancels $D(\sigma)$,
which is (ii). For (iii), $\sum_{j}x_j=M$ at every profile by
Lemma~\ref{lem:clear}(i), so $\sum_{j}[x_j(\sigma)-x_j(\mathbf{0})]=0$;
splitting the sum over $\mathcal{U}$ and $\mathcal{C}$ gives
\eqref{eq:conserve}.
\end{proof}

The result holds at every profile, so an event can lose visibility without
being targeted: all bystanders change by the same factor, preserving their
relative order, and their aggregate loss equals the contested events' gain.

\subsection{Memory capture}
\vspace{-1mm}
Let $C\subset\NN$ coordinate its members' preservation efforts over a
nonempty target set $\E_C\subsetneq\E$, and write $\bar B_C$ for the total
preservation effort assigned to that set, not the members' combined budget.
Since outside agents may suppress the targets, we bound that counter-effort
and the outside scores.

\begin{theorem}[Coordinated memory-capture guarantee]\label{thm:capture}
Fix $\varepsilon\in(0,1)$ and $\lambda>0$, and suppose that
(a) every $i\in C$ satisfies $\E_C\subseteq\E_i$, so each member is
eligible to act on every target event, and
(b) the prescribed effort is \emph{individually feasible}: there exist
$b_i\ge0$ with $\sum_{i\in C}b_i=\bar B_C$ such that agent $i$ can spread
$b_i$ evenly over $\E_C$ within its own budget, $T_i^\tau\le B_i$; at
$\tau=0$ this holds if and only if
$\bar B_C\le\sum_{i\in C}B_i/c_r$.
The coalition uses preservation
effort only, spread evenly over $\E_C$, and the remaining agents satisfy
\begin{equation}\label{eq:oppbounds}
S_k\le\bar S\ \ \text{for } k\notin\E_C,
\qquad
\sum_{i\notin C}f_i(e)\le\bar F\ \ \text{for } e\in\E_C .
\end{equation}
Define
\begin{equation}\label{eq:bstar}
B^\star:=|\E_C|\Big[\bar S-\min_{e\in\E_C}\omega_e+\bar F
+\tfrac1\lambda\ln\tfrac{(1-\varepsilon)|\E\setminus\E_C|}
{\varepsilon|\E_C|}\Big].
\end{equation}
If $B^\star>0$ and $\bar B_C\ge B^\star$ then
$\sum_{e\in\E_C}x_e\ge(1-\varepsilon)M$ at every such profile; if
$B^\star\le0$ the same holds at $\bar B_C=0$.
\end{theorem}

\begin{proof}
Put $T=\sum_{e\in\E_C}e^{\lambda S_e}$ and
$V=\sum_{k\notin\E_C}e^{\lambda S_k}$. By \eqref{eq:oppbounds},
$S_e\ge\underline S:=\min_{e\in\E_C}\omega_e+\bar B_C/|\E_C|-\bar F$ for
$e\in\E_C$ while $S_k\le\bar S$ otherwise, so
$T\ge|\E_C|e^{\lambda\underline S}$ and
$V\le|\E\setminus\E_C|e^{\lambda\bar S}$. Since
$\sum_{e\in\E_C}x_e=MT/(T+V)$ increases in $T$ and decreases in $V$, it is
at least $(1-\varepsilon)M$ as soon as
$\underline S\ge\bar S+\lambda^{-1}
\ln[(1-\varepsilon)|\E\setminus\E_C|/(\varepsilon|\E_C|)]$. Substituting
$\underline S$ gives $\bar B_C\ge B^\star$, sufficient when $B^\star>0$;
if $B^\star\le0$ the inequality holds at $\bar B_C=0$.
\end{proof}

Theorem~\ref{thm:capture} is a sufficient allocation guarantee for the
prescribed equal split, against the stated bounds. Hypothesis (b) is what
makes the aggregate bound $\sum_{i\in C}B_i/c_r$ usable: eligibility alone
does not ensure the split can be financed within individual budgets. The
theorem does not show that coordination is individually optimal, that the
profile is a Nash equilibrium, or that the group is stable, and it does not
treat endogenous coalition formation. At $\tau=0$,
$\bar F=\sum_{i\notin C}B_i/c_f$ and
$\bar S=\max_{k\notin\E_C}\omega_k+\sum_{i\notin C}B_i/c_r$. When
$\varepsilon<|\E\setminus\E_C|/|\E|$ the threshold falls in $\lambda$ and
rises in $\bar F$ at rate $|\E_C|$; outside that range the logarithm in
\eqref{eq:bstar} changes sign and the comparative static reverses. The
threshold is independent of $M$: securing a given capacity \emph{share} does
not depend on total capacity.

\section{THE PLATFORM'S DESIGN PROBLEM}
\label{sec:mech}
\vspace{-1mm}
Effort can change visibility even when the record does not change. The
allocation depends on the clearing gain $\lambda$ and the cost curvature
$\tau$, which we compare as design choices: $\lambda$ sets how strongly score
differences affect allocation, $\tau$ the shape of the cost schedule in
\eqref{eq:budget} rather than a fixed fee. The exercise treats alternative
cost schedules as available parameter choices; it does not assume the
platform can change agents' endowments. Convex costs also limit effort
independently of the budget.

\begin{remark}[Budget-free saturation]\label{rem:sat}
Let $\tau>0$ in a restricted contest under \textbf{(A3)}--\textbf{(A4)}.
The payoff derivative in an agent's own coordinate is
$g_i\lambda x_e(1-x_e/M)-c(1+\tau)y^\tau$, whose first term is at most
$g_i\lambda M/4$ by Lemma~\ref{lem:clear}(iv), so every best response
there obeys $y_i^\star\le\big(g_i\lambda M/[4c(1+\tau)]\big)^{1/\tau}$,
a cap involving neither $B_i$ nor, being a single-agent bound, the
aggregate effort of a coalition.
\end{remark}

\subsection{The design problem}
\vspace{-1mm}
Assume $\sum_{k\in\E}\omega_k>0$ and let
$x^{\mathrm{ref}}_e:=M\omega_e/\sum_{k\in\E}\omega_k$ be the policy target: a
normative benchmark representing a mandate to allocate attention in
proportion to independently assessed importance, not something implied by the
softmax rule. Let $\mathcal S(\lambda,\tau)$ be the set of pure-strategy
MMEs, when nonempty. Since the analysis does not establish that this set is a
singleton, or even nonempty, for every pair, we use an explicit numerical
selection rule: $\widehat\sigma(\lambda,\tau)$ is the equilibrium returned by
iterated global best response from zero effort. Runs from other feasible
starts serve as robustness checks and do not establish uniqueness.

The platform-design comparison is carried out over a finite set
$\mathcal G\subset\Lambda\times\mathrm T$ of evaluated design pairs:
\begin{equation}\label{eq:design}
\begin{aligned}
\min_{(\lambda,\tau)\in\mathcal G}\quad
& J(\lambda,\tau):=\frac1{2M}\sum_{e\in\E}
\big|x_e(\widehat\sigma(\lambda,\tau))-x^{\mathrm{ref}}_e\big| \\
\text{s.t.}\quad & x_e(\widehat\sigma(\lambda,\tau))\ \ge\ \underline x_e,
\qquad e\in\E_{\mathrm{crit}},
\end{aligned}
\end{equation}
Here $\mathcal G$ is the finite grid evaluated numerically and
$\E_{\mathrm{crit}}$ the events with visibility floors $\underline x_e$. The
reported design is the best feasible evaluated pair under that selection
rule, not a continuous optimum over the parameter region nor an optimum over
all equilibria, and no monotonicity of $J$ is claimed.
Theorem~\ref{thm:mono} gives uniqueness at $\tau=1$ only where its curvature
condition holds; elsewhere the comparison uses the selected equilibrium.

We use two references: the proportional-importance target, and the no-effort
allocation $x^0(\lambda)$ generated by the rule itself. Distance from the
first measures total mismatch; distance from the second measures what
strategic effort adds. They need not sum to the total, since they can partly
offset one another, and a gap between $x^0(\lambda)$ and the proportional
target is a feature of the rule rather than evidence of manipulation. With
$x^{0}_e(\lambda)=Me^{\lambda\omega_e}/\sum_ke^{\lambda\omega_k}$, put
\begin{equation}\label{eq:split}
\begin{aligned}
D^{\mathrm{base}}(\lambda)&:=\tfrac1{2M}\textstyle\sum_{e}
 |x^{0}_e(\lambda)-x^{\mathrm{ref}}_e| ,\\
D^{\mathrm{str}}(\lambda,\tau)&:=\tfrac1{2M}\textstyle\sum_{e}
 |x_e(\widehat\sigma(\lambda,\tau))-x^{0}_e(\lambda)| .
\end{aligned}
\end{equation}

\begin{proposition}[Benchmark decomposition]\label{prop:split}
For every $(\lambda,\tau)$ at which the selected profile $\widehat\sigma$ is defined,
$|D^{\mathrm{base}}-D^{\mathrm{str}}|\le J\le
D^{\mathrm{base}}+D^{\mathrm{str}}$. The term $D^{\mathrm{base}}$ is
free of $\tau$ and of every agent's behaviour, and if the $\omega_e$ take
at least three distinct positive values then
$D^{\mathrm{base}}(\lambda)>0$ for every $\lambda>0$.
\end{proposition}

\begin{proof}
Both inequalities are the $\ell_1$ triangle inequality applied to
$x(\widehat\sigma)-x^{\mathrm{ref}}=(x(\widehat\sigma)-x^{0})+
(x^{0}-x^{\mathrm{ref}})$, divided by $2M$. For the last claim,
$D^{\mathrm{base}}=0$ would need $h(t):=e^{\lambda t}/t$ to take the same
value at every $\omega_e$; as $h'(t)=e^{\lambda t}(\lambda t-1)/t^2$, $h$
strictly decreases then strictly increases and so attains each value at
most twice, which three distinct levels cannot satisfy.
\end{proof}

\noindent The softmax-consistent benchmark is $x^0(\lambda)$, for which the
baseline displacement is zero by definition; we therefore report
$D^{\mathrm{str}}$ alongside the total mismatch $J$ from the normative
target.

\begin{theorem}[Feasibility floor as $\tau\to\infty$]\label{thm:floor}
Fix all parameters except $\tau$. Then:
\begin{enumerate}
  \item every effort coordinate satisfies
        $\sigma^\star_{i\ell}(\tau)\le (B_i/c_\ell)^{1/(1+\tau)}\to 1$,
        a bound inherited from feasibility alone;
  \item consequently $\liminf_{\tau\to\infty} J(\lambda,\tau)\ge
        \min_{\sigma\in[0,1]^d}\mathcal D_\lambda(\sigma)$, where
        $\mathcal D_\lambda(\sigma):=(2M)^{-1}\sum_{e}|x_e(\sigma)-x_e^{\mathrm{ref}}|$;
  \item this relaxation-based floor is strictly positive if some
        $\omega_e=0$, or if
$|\lambda^{-1}\ln(\omega_e/\omega_k)-(\omega_e-\omega_k)|>2n$
        for some $e,k\in\E$.
\end{enumerate}
\end{theorem}
\begin{proof}
(1) Every term of the budget constraint~\eqref{eq:budget} is nonnegative,
so each coordinate satisfies $c_\ell\,\sigma_{i\ell}^{1+\tau}\le B_i$,
which gives $\sigma_{i\ell}\le (B_i/c_\ell)^{1/(1+\tau)}$. The right-hand
side tends to $1$ as $\tau\to\infty$. This is a coordinate-wise
consequence of the budget constraint and uses no equilibrium condition;
in particular it does not assert that equilibrium efforts converge.

(2) Take a sequence $\tau_n\to\infty$ along which $J(\lambda,\tau_n)$
converges to $\liminf_{\tau\to\infty}J(\lambda,\tau)$. By~(1) the profiles
$\sigma^\star(\tau_n)$ are eventually bounded, so a subsequence converges
to some $\bar\sigma$, and by~(1) again $\bar\sigma\in[0,1]^d$: the unit
box is the limit of the feasible boxes, not a set of equilibria. Since
$\mathcal D_\lambda$ is continuous,
$\liminf_{\tau\to\infty}J(\lambda,\tau)=\mathcal D_\lambda(\bar\sigma)
\ge\min_{\sigma\in[0,1]^d}\mathcal D_\lambda(\sigma)$. The right-hand side
is a relaxation: it minimizes over the whole box, not over the equilibria
the box contains.

(3) The minimum in~(2) is attained, because $[0,1]^d$ is compact and
$\mathcal D_\lambda$ is continuous. It is therefore positive unless some
$\sigma\in[0,1]^d$ gives $x(\sigma)=x^{\mathrm{ref}}$. If $\omega_e=0$,
then $x_e^{\mathrm{ref}}=0$, whereas $x_e(\sigma)>0$ for every $\sigma$ by
Lemma~\ref{lem:clear}(i). If all $\omega_e>0$, then by
\eqref{eq:clear} the equality
$x(\sigma)=x^{\mathrm{ref}}$ requires
$S_e-S_k=\lambda^{-1}\ln(\omega_e/\omega_k)$ for all $e,k$, that is,
$\nu_e-\nu_k=\lambda^{-1}\ln(\omega_e/\omega_k)-(\omega_e-\omega_k)$.
On $[0,1]^d$ each net effort satisfies $\nu_e\in[-n,n]$, so
$|\nu_e-\nu_k|\le 2n$, which the condition in~(3) rules out for some
pair. In both cases no profile in the box reproduces $x^{\mathrm{ref}}$,
and the minimum is strictly positive.
\end{proof}
Theorem~\ref{thm:floor} shows that costly effort alone cannot make memory
accurate: steep costs cap effort, but the softmax responds to score
differences while the benchmark follows importance ratios, and bounded effort
cannot always close the gap. Part~(3) is sufficient, not necessary --- its
failure does not imply a zero floor. By Proposition~\ref{prop:split}, raising
$\tau$ acts only on $D^{\mathrm{str}}$, which does not make $J$ monotone in
$\tau$; removing the floor requires changing the clearing rule itself.

%==============================================================================
\begin{figure*}[!t]
\centering
\setlength{\tabcolsep}{2pt}
\begin{tabular}{cccc}
\begin{minipage}[t]{0.19\textwidth}\centering
  \includegraphics[width=\linewidth]{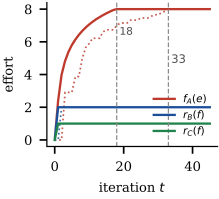}\par
  \vspace{-1mm}{\scriptsize (a) Allocation dynamics}
\end{minipage} &
\begin{minipage}[t]{0.19\textwidth}\centering
  \includegraphics[width=\linewidth]{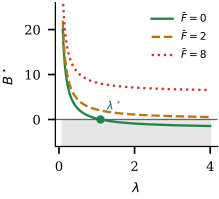}\par
  \vspace{-1mm}{\scriptsize (b) Capture threshold}
\end{minipage} &
\begin{minipage}[t]{0.19\textwidth}\centering
  \includegraphics[width=\linewidth]{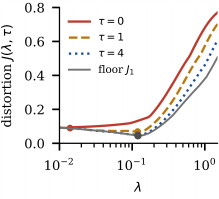}\par
  \vspace{-1mm}{\scriptsize (c) Platform design}
\end{minipage} &
\begin{minipage}[t]{0.19\textwidth}\centering
  \includegraphics[width=\linewidth]{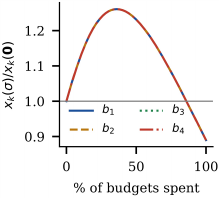}\par
  \vspace{-1mm}{\scriptsize (d) Bystander scaling}
\end{minipage}
\end{tabular}
\caption{(a) Adjustment trajectories and resulting allocation in the
two-event contest; (b) capture effort under different outside-score
bounds; (c) total distortion $J$ for $\tau=0,1,4$, with baseline term
$D^{\mathrm{base}}$ (dashed) and strategic term $D^{\mathrm{str}}$ at
$\tau=0$ (dotted); (d) common scaling of bystander allocations.}
\label{fig:panels}
\end{figure*}

\section{NUMERICAL OBSERVATIONS}
\label{sec:num}
\vspace{-2mm}
Unless stated otherwise the six-event instance has $\omega=(10,8,9,7,6,5)$,
$M=100$, $\lambda=0.5$ and unit effort costs. Agent $A$ is named in $e_1$
with $\theta_{e_1}=-1$, so by \eqref{eq:stake} with $\rho=5$ its stake is
$\gamma_A(e_1)=-5$: it gains from lower visibility and suppresses $e_1$, with
budget $8$. Agents $B$ and $C$ are named in $e_2$ with $\theta_{e_2}=+1$, so
$\gamma_B(e_2)=\gamma_C(e_2)=+5$ and both preserve $e_2$, with budgets $2$ and
$1$; all $\beta_{ik}=0$. With $\E_A=\{e_1\}$ and $\E_B=\E_C=\{e_2\}$ only
$e_1$ and $e_2$ receive effort and the other four are bystanders. These
eligibility restrictions are part of this instance.

\emph{Computation and verification.} Best responses are found by global
maximization over the feasible interval, including endpoints and interior
stationary points, since payoffs can be non-concave at $\tau=0$. We select
the equilibrium reached from zero effort, repeat from ten random feasible
starts, and check each profile by re-maximizing every agent's payoff against
the others. These checks support the reported equilibria but do not establish
uniqueness, which is claimed only where Theorem~\ref{thm:mono} applies.

\textit{Overlapping eligibility and rivalry:}
To test cross-event coupling, take $\omega=(10,8,6)$, $M=100$,
$\lambda=0.015$, $\tau=1$, unit costs and budgets $B_A=B_B=4$.
Both agents can act on $e_1,e_2$; $e_1$ names $A$ and $e_2$ names $B$.
With $\rho=5$, $\theta_{e_1}=\theta_{e_2}=1$,
$\beta_{AB}=0.6$ and $\beta_{BA}=0.8$, equation \eqref{eq:stake} gives
$\gamma_A=(\rho,-\rho\beta_{AB})=(5,-3)$ and
$\gamma_B=(-\rho\beta_{BA},\rho)=(-4,5)$, the rivalry terms being
$\rho\beta_{ik}$ rather than $\beta_{ik}$. Thus both agents have stakes in
both contested events and both rivalry coefficients are positive.
Here $\hat g=5$ and $L=1.125<2c_{\min}=2$, so Theorem~\ref{thm:mono}
applies. With $\eta=0.05$, projected-gradient play reaches
$(r_A(e_1),f_A(e_2),f_B(e_1),r_B(e_2))
=(1.1028,0.9302,1.1056,1.1779)$ and allocation
$(34.295,33.406,32.299)$. The agents spend $2.0813$ and
$2.6100$ of their budgets of $4$, so neither constraint binds. Ten starts
over the four coordinates reach the same profile to numerical tolerance.

\textit{Allocation and equilibrium:}
In the six-event instance, the equilibrium allocation is
$(0.675,60.746,22.347,8.221,4.986,3.024)$: event $e_1$ has the highest
importance but the lowest share, because $A$ spends its whole budget
suppressing it. The four bystanders
retain their scores and are all rescaled by $0.8898$; their combined
loss, $4.779$ units, equals the gain of the two contested events by
Proposition~\ref{prop:crowd}. The profile is a pure MME by
Corollary~\ref{cor:corner}. Over the full joint feasible set, the
minimum of $x_{e_1}(1-x_{e_1}/M)$ is $0.6703$, and that of
$x_{e_2}(1-x_{e_2}/M)$ is $12.9128$. The corresponding marginal
benefits are $1.676$, $32.282$ and $32.282$, all above the unit
effort cost. Hence full-budget effort is a strict best response for
each agent against every feasible choice of the others.

\textit{Dynamics and the certified region:}
For $\tau=1$ and unit costs, Theorem~\ref{thm:mono} certifies
$\lambda\le0.0163$. At $\lambda=0.016$, $\mu=0.08$ and $\ell=3.92$, so
$\eta<0.0104$; with $\eta=0.01$ the iteration reaches error $10^{-6}$ in
$688$ steps, against the certified modulus $q=0.99997$ of \eqref{eq:rate} and
an observed per-step contraction of $0.980$ --- a worst-case bound against an
empirical rate on this instance. At $\lambda=0.117$, outside the certified
region, convergence takes $57$ iterations at observed rate $0.364$, with no
guarantee. Ten starts converge to the same limit at all gains tested up to
$1$. In the two-event $\tau=0$ example, exact and asynchronous/inexact
updates reach the full-budget profile in $18$ and $33$ iterations; the latter
is empirical and outside the scope of Theorem~\ref{thm:mono}.

\textit{Coalition capture:}
Panel (b) uses $\varepsilon=0.1$, $|\E_C|=|\E\setminus\E_C|=1$, minimum
target importance $10$ and outside-score bound $8$. With no
counter-effort, \eqref{eq:bstar} reduces to
$B^\star=\ln(9)/\lambda-2$, which is zero at $\lambda=1.0986$;
counter-effort bounds $2$ and $8$ add $2|\E_C|$ and $8|\E_C|$. Since
$\varepsilon=0.1<|\E\setminus\E_C|/|\E|=0.5$, the threshold is decreasing
in $\lambda$ over this range.
This is a worst-case sufficient guarantee under equal splitting.

\textit{Platform design and bystander scaling:}
Panel (c) gives the best evaluated designs against the proportional-
importance target: $(\lambda,\tau)=(0.014,0)$, $(0.117,1)$ and
$(0.120,4)$, with $J=0.0895$, $0.0691$ and $0.0488$. The latter
two satisfy $x_{e_1}\ge15$, with allocations $15.80$ and $18.75$.
Table~\ref{tab:split} separates baseline mismatch from strategic
displacement. At $\lambda=0.014$ the selected profile is zero effort,
since the marginal benefit $g\lambda x_{e_1}(1-x_{e_1}/M)=0.999$ falls
just below the unit cost and Lemma~\ref{lem:corner}(ii) applies, yet the
baseline mismatch is $0.0895$. The baseline mismatch is smallest near
$\lambda=0.117$; at this gain, increasing $\tau$ from $1$ to $4$
reduces $D^{\mathrm{str}}$ from $0.0609$ to $0.0461$.  Panel (d) shows the non-monotone bystander factor:
it rises to $1.2609$ at $36.1\%$ budget use and falls to $0.8898$
at full expenditure.

\begin{table}[!t]\centering\footnotesize
\setlength{\tabcolsep}{3pt}
\caption{Best evaluated designs and baseline gain. $J$ is total mismatch; $D^{\mathrm{base}}$ and $D^{\mathrm{str}}$ are its baseline and strategic components. Efforts are $(f_A(e_1),r_B(e_2),r_C(e_2))$.}
\label{tab:split}
\vspace{-1mm}
\begin{tabular}{ccccc l}
\toprule
$\tau$ & $\lambda$ & $J$ & $D^{\mathrm{base}}$ & $D^{\mathrm{str}}$ & Effort\\
\midrule
$0$ & $0.014$ & $0.0895$ & $0.0895$ & $0.0000$ & $(0,0,0)$\\
$1$ & $0.117$ & $0.0691$ & $0.0146$ & $0.0609$ & $(2.83,1.41,1)$\\
$4$ & $0.120$ & $0.0488$ & $0.0128$ & $0.0461$ & $(1.16,1.15,1)$\\
$0$ & $0.500$ & $0.4532$ & $0.2430$ & $0.4551$ & $(8,2,1)$\\
\bottomrule
\end{tabular}
\end{table}
\vspace{-1mm}
\section{CONCLUSIONS}
\label{sec:concl}
\vspace{-2mm}
We studied a game in which agents use budgeted preservation and suppression efforts to move the visibility of past events under a fixed-capacity softmax rule. We establish mixed-strategy existence and a sufficient condition for pure-strategy existence; under quadratic costs, an explicit curvature bound gives equilibrium uniqueness and geometric convergence of synchronous projected-gradient play. Strategic effort can reverse visibility rankings and rescale events receiving no direct effort, and a prescribed coordinated policy attains a specified memory share under a sufficient condition. Numerical comparisons separate the no-effort mismatch from the displacement strategic effort adds, restricted to the evaluated grid and the stated selection rule. Endogenous coalition formation and reader welfare are interesting future directions.

%------------------------------------------------------------------
% Reference list (24 entries), ECC / IEEE CSS style. UNCHANGED from v21
% except: Tullock1980 removed with the rent-dissipation remark, and the
% Dellarocas year restored.
%------------------------------------------------------------------

\end{document}